\documentstyle[texdraw,12pt,amsmath,a4]{article}

\begin{document}

\begin{flushright}
SHEP 99-07\\ 
hep-lat/9906031
\end{flushright}
\vspace{1cm}

\begin{center}
{\Large \bf
An Exploratory Lattice Study of Spectator Effects in Inclusive
Decays of the $\mathbf{\Lambda_b}$ Baryon}
\vskip 5mm
{ \bf UKQCD Collaboration}: \\ \vskip 2mm 
Massimo Di Pierro and Chris T Sachrajda\\ 
Department of Physics and Astronomy \\ 
University of Southampton, SO17 1BJ, UK \\
\vskip 2mm
and \\ 
\vskip 2mm
Chris Michael\\ 
Theoretical Physics Division\\  
Dept. of Math. Sciences 
\\ University of Liverpool, L69 3BX, UK
\end{center}

\begin{abstract}
A possible explanation of the apparent disrepency between the
theoretical prediction and experimental measurement of the ratio of
lifetimes $\tau(\Lambda_b)/\tau(B_d)$ is that ``spectator effects'',
which appear at $O(1/m_b^3)$ in the heavy quark expansion, contribute
significantly. We investigate this possibility by computing the
corresponding operator matrix elements in a lattice simulation. We
find that spectator effects are indeed significant, but do not appear
to be sufficiently large to account for the full discrepency. We
stress, however, that this is an exploratory study, and it is
important to check our conclusions on a larger lattice and using a
larger sample of gluon configurations. \\ 

\end{abstract}

\section{Introduction}
At leading order in the heavy-quark expansion the decay rate of the
heavy quark is independent of its parent hadron. In this letter we
present the results of an exploratory study, in which we attempt to gain
some understanding of the striking discrepancy between the experimental
result for the ratio of lifetimes~\cite{junk}
\begin{equation}
\frac{\tau(\Lambda_b)}{\tau(B_d)}=0.78 \pm 0.04
\label{exp1}
\end{equation}
and the theoretical prediction~\cite{ns}, based on the heavy-quark
expansion~\cite{chay}--\cite{reviews2}
\begin{equation}
\frac{\tau(\Lambda_b)}{\tau(B_d)}=0.98+O(1/m_b^3)\ .
\label{theo1}
\end{equation}
In particular, we compute the contributions to the $O(1/m_b^3)$ term on
the right-hand side of eq.~(\ref{theo1}) which come from ``spectator
effects'', i.e. from decays in which two quark or antiquark constituents
of the beauty hadron participate in the weak decay. These effects may be
larger than estimates based purely on power counting would indicate as
a result of the enhancement of the phase space for $2 \rightarrow 2$
body reactions relative to $1 \rightarrow 3$ body decays~\cite{ns}. 

Denoting the $O(1/m_b^3)$ contribution from spectator effects by
$\Delta$ and following the analysis of ref.~\cite{ns} we find that
\footnote{ Equation (\ref{expansion1}) can be derived from eq.(39) of
ref.~\cite{ns} by the substitution
\begin{eqnarray}
\widetilde{B} &=&-6L_1 \label{eq:btildedef}\\
r &=&-2\frac{L_2}{L_1}-\frac 13\ .\label{eq:rdef}
\end{eqnarray}}
\begin{equation}
\Delta=-0.173 \varepsilon_1 +0.195 \varepsilon_2 +0.030
L_1 - 0.252 L_2
\label{expansion1}
\end{equation}
where
\begin{eqnarray}
\varepsilon _1(m_b) &=&\frac 8{f_B^2m_B}\frac{\left\langle B_q\right| 
\overline{b}\gamma ^\mu Lt^aq\overline{q}\gamma ^\mu Lt^ab\left|
B_q\right\rangle }{2m_B}=-0.01\pm 0.03 \ ,\\
\varepsilon _2(m_b) &=&\frac 8{f_B^2m_B}\frac{\left\langle B_q\right| 
\overline{b}Lt^aq\overline{q}Rt^ab\left| B_q\right\rangle }{2m_B}=-0.02\pm
0.02
\end{eqnarray}
have been computed, using lattice simulations, in ref.~\cite{me1}. 
The two new variables introduced here are
\begin{eqnarray}
L_1(m_b) &=&\frac 8{f_B^2m_B} \frac{\left\langle \Lambda _b
\right| \overline{b}\gamma ^\mu Lq\overline{q}\gamma ^\mu Lb\left|
\Lambda _b \right\rangle }{2m_{\Lambda _b}} 
\label{elle1} \\
L_2(m_b) &=&\frac 8{f_B^2m_B}\frac{\left\langle \Lambda _b
\right| \overline{b}\gamma ^\mu Lt^aq\overline{q}\gamma ^\mu
Lt^ab\left| \Lambda_b \right\rangle }{2m_{\Lambda _b}}\ . 
\label{elle2}
\end{eqnarray}
Heavy-quark symmetry implies that there are only two matrix elements
which need to be considered for the $\Lambda_b$, in contrast to the
four for $B$-mesons~\cite{ns}~\footnote{The coefficients of two of
the matrix elements for $B$-mesons are so small that we don't include them
in eq.~(\ref{expansion1}).}.

We stress that this is an exploratory study. It is the first calculation
of the matrix elements $L_{1,2}$ and provides a preliminary indication
of whether spectator effects can reconcile the experimental measurements
and theoretical predictions for the ratios of lifetimes in
eqs.~(\ref{exp1}) and (\ref{theo1}). We perform the calculations with a
static $b$-quark and two (rather large) values of the mass of the
light-quark and do not attempt to extrapolate the results to the chiral
limit. Our results indicate that spectator effects are not negligible,
although they do not appear to be sufficiently large to account fully
for the discrepency. Specifically we find:
\begin{eqnarray}
L_1(m_b)&=&\left\{
	\begin{tabular}{ll}
	$-0.31(3)$ & for $a m_\pi=0.74(4)$\\
	$-0.22(4)$ & for $a m_\pi=0.52(3)\ ,$
	\end{tabular}
	\right. \\ 
L_2(m_b)&=&\left\{
	\begin{tabular}{ll}
	$0.23(2)$ & for $a m_\pi=0.74(4)$\\
	$0.17(2)$ & for $a m_\pi=0.52(3)\ ,$
	\end{tabular}
	\right. 
\end{eqnarray}
with $a^{-1}\simeq 1.1$\, GeV. The corresponding results for the ratio
of lifetimes are
\begin{equation}
\frac{\tau(\Lambda_b)}{\tau(B_d)}=	\left\{
	\begin{tabular}{ll}
	$0.91(1)$ & for $a m_\pi=0.74(4)$\\
	$0.93(1)$ & for $a m_\pi=0.52(3)$\ .
	\end{tabular}
	\right. 
\label{theo2}
\end{equation}

\section{Perturbative matching}

In this section we briefly discuss the matching factors which are
required to obtain the matrix elements of the continuum four-quark
operators renormalised at a scale $\mu$ from those of the bare lattice
operators computed in lattice simulations at a cut-off $a^{-1}$. The
details of the calculation are presented in ref.~\cite{me1}. Here we
simply summarise the main points required for the evaluation of the
matrix elements $L_1$ and $L_2$ in the $\overline{\textrm{MS}}$ scheme.

We start by using the renormalisation group to relate the matrix
elements $L_1$ and $L_2$, defined in the $\overline{\text{MS}}$ scheme
at two different renormalisation scales, $\mu=m_b$ and $\mu =
a^{-1}$. Since the Wilson coefficient functions in the OPE expansion
(\ref{expansion1}) have been evaluated at tree level only~\cite{ns},
we keep just the leading logarithms in the evolution equations so
that~\cite{shif1, shif2}
\begin{equation}
\binom{L_1(m_b)}{L_2(m_b)}=\left( 
\begin{array}{ll}
1+\frac{2C_F\delta }{N_c} & -\frac{2\delta }{N_c} \\ 
-\frac{C_F\delta }{N_c^2} & 1+\frac \delta {N_c^2}
\end{array}
\right) \binom{L_1(a^{-1})}{L_2(a^{-1})}\ ,
\end{equation}
where
\begin{equation}
\delta =\left( \frac{\alpha _s^{\overline{\text{MS}}}(a^{-1})}{\alpha
_s^{\overline{\text{MS}}}(m_b)} \right)^{9/2\beta_0}-1=0.40 \pm 0.04\ .
\end{equation}
In estimating $\delta$ we have used $\Lambda_{\text{QCD}}=250$~MeV,
$a^{-1}=1.10$~GeV, $m_b=4.5$~GeV and $\beta_0=9$. The error in
$\delta$ is evaluated includes a 20\% uncertainty for
$\Lambda_{\text{QCD}}$.

In the second step of the matching we relate the matrix elements
renormalised in the continuum to those regularized on lattice, both at
the same scale, $a^{-1}$. Although this involves corrections of
$O(\alpha_s)$, which are, in principle, beyond the precision which we
require, we nevertheless include them because the perturbative
coefficients in lattice perturbation theory are generally large. The
computation for the most general four quark operator involving one
heavy quark appears in the appendix of ref.~\cite{me1}. For $L_1$
and $L_2$ the relevant relations are:
\begin{eqnarray}
L_1(a^{-1})& = & \frac{8}{f_B^2m_B} \left[ h_{11} M_1 + h_{12} M_2 +
h_{13} M_3 + h_{14} M_4 \right] \\ 
L_2(a^{-1})& = &\frac{8}{f_B^2m_B} \left[ h_{21} M_1 + h_{22} M_2 +
h_{23} M_3 + h_{24} M_4 \right]\ ,
\end{eqnarray}
where 
\begin{eqnarray}
M_1 &=&\frac{\left\langle \Lambda _b\right| (\overline{b}\gamma ^\mu Lq)
\ (\overline{q}\gamma ^\mu Lb)\left| \Lambda _b\right\rangle }
{2m_{\Lambda _b}}\label{eq:m1}\\
M_2 &=&\frac{\left\langle \Lambda _b\right| (\overline{b}\gamma ^\mu
L\gamma ^0q)\ (\overline{q}\gamma ^\mu Lb)\left| \Lambda _b\right\rangle}
{2m_{\Lambda _b}}+
\frac{\left\langle \Lambda _b\right| (\overline{b}\gamma ^\mu Lq)\ 
(\overline{q}
\gamma ^0\gamma ^\mu Lb)\left| \Lambda _b\right\rangle }{2m_{\Lambda _b}}
\label{eq:m2}\\
M_3 &=&\frac{\left\langle \Lambda _b\right| (\overline{b}\gamma ^\mu Lt^aq)
\ (\overline{q}\gamma ^\mu Lt^ab)\left| \Lambda _b\right\rangle }
{2m_{\Lambda _b}} \label{eq:m3}\\
M_4 &=&\frac{\left\langle \Lambda _b\right| (\overline{b}\gamma ^\mu L\gamma
^0t^aq)(\overline{q}\gamma ^\mu Lt^ab)\left| \Lambda _b\right\rangle }
{2m_{\Lambda _b}}+\frac{\left\langle \Lambda _b\right| (\overline{b}\gamma
^\mu Lt^aq)(\overline{q}\gamma ^0\gamma ^\mu Lt^ab)\left| \Lambda
_b\right\rangle }{2m_{\Lambda _b}}
\label{eq:m4}
\end{eqnarray}
are the matrix elements of the bare lattice operators regularised at
$a^{-1}$. The coefficients $h_{ij}$ are listed in table~(\ref{table1})
where for the lattice coupling constant we have used a boosted coupling
equal to
\begin{equation}
\frac{\alpha _s^{\text{latt}}(a^{-1})}{4\pi }=\frac{6(8 \kappa_{crit})^4}{(4
\pi)^2 \beta}\simeq 0.01216\ .
\end{equation}
Readers who prefer to use other choices of the lattice coupling constant
can combine the coefficients in table~\ref{table1} with their choice
of coupling. 

\begin{table}
\begin{center}
\setlength{\tabcolsep}{0.5pc}
\newlength{\digitwidth} \settowidth{\digitwidth}{\rm 0}
\catcode`?=\active \def?{\kern\digitwidth}
\begin{tabular}{|c|c|c|} \hline
coeff. & expression & value \\ \hline
$h_{11}$ & $1+\frac \alpha {4\pi }\left[ 
	\frac{10}3-\frac 43x_1-\frac 83x_2
	\right]\simeq 1-21.65\frac\alpha {4\pi}$ & $0.737$ \\ 
$h_{12}$ & $\frac \alpha {4\pi }\left[ 
	-\frac 43x_3
	\right]\simeq 9.19\frac\alpha {4\pi}$ 
 & $0.112 $\\ 
$h_{13}$ & $\frac \alpha {4\pi }\left[ 
	-\frac 52-x_4-2x_5-x_7
	\right]
\simeq 9.29\frac\alpha {4\pi}$ 
& $0.113$ \\ 
$h_{14}$ & $\frac \alpha {4\pi }\left[ 
	-x_6
	\right] 
\simeq 6.89\frac\alpha {4\pi}$ 
& $0.084$ \\ 
$h_{21}$ & $\frac \alpha {4\pi }\frac29\left[ 
	-\frac 52-x_4-2x_5-x_7
	\right]
\simeq 2.06\frac\alpha {4\pi}$ 
&$ 0.025$ \\ 
$h_{22}$ & $\frac \alpha {4\pi }\left[ 
	-\frac 29x_6
	\right]
\simeq 1.53\frac\alpha {4\pi}$ 
& $0.019 $\\ 
$h_{23}$ & $1+\frac \alpha {4\pi }\left[ 
	\frac5{12}-\frac 43x_1+\frac 13x_2
	-\frac76x_4+\frac 23x_5-\frac 76x_7
	\right]
\simeq 1-10.82\frac\alpha {4\pi}$ 
& $0.869$ \\ 
$h_{24}$ & $\frac \alpha {4\pi }\left[ 
	\frac12 x_6
	\right]
\simeq -3.45\frac\alpha {4\pi}$ 
& $-0.042$ \\ \hline
\end{tabular}
\end{center}
\caption{Matching coefficients for the matrix elements $M_1$ and $M_2$,
	renormalised on lattice at an energy scale $a^{-1}=1.10$ GeV. 
	The values of the integrals $x_i$ are reported in the Appendices
	of ref.~\cite{me1}.\label{table1}}
\end{table}

A consequence of the Heavy Quark Effective Theory, and the fact that the
light quarks in the $\Lambda_b$ are in a spin zero combination, is that
the number of lattice operators whose matrix elements have to be
evaluated is four rather than 8 (which is the case for heavy
mesons~\cite{me1}).

In order to obtain the factor $f_B^2 m_B$ it is also
necessary to determine the normalization of the axial current, $Z_A$,
\begin{equation}
f_B m_B = \left<0\right| A_0(a^{-1}) \left| B \right> 
= Z_A \left<0\right| A_0^{\text{latt}}(a^{-1}) \left| B \right> = \sqrt{2
m_B} Z_A Z_L\ ,
\end{equation}
where $A_0^{\text{latt}}$ is the time component of the axial current
defined on the lattice and $Z_L$ is obtained from the matrix element
determined in numerical simulations At our value of the lattice
spacing
\begin{equation}
Z_A=1-20.0\frac{\alpha_s^{\text{latt}}(a^{-1})}{4 \pi} \simeq 0.75\ .
\end{equation}
In the following section we combine the results for the matrix
elements computed on the lattice ($M_i$ and $Z_L$) with the perturbative
coefficients presented in this section ($h_{ij}$ and $Z_A$), to obtain
the values of $L_1(m_b)$ and $L_2(m_b)$.
   
\section{Lattice computation and results}
\begin{figure}
	\input{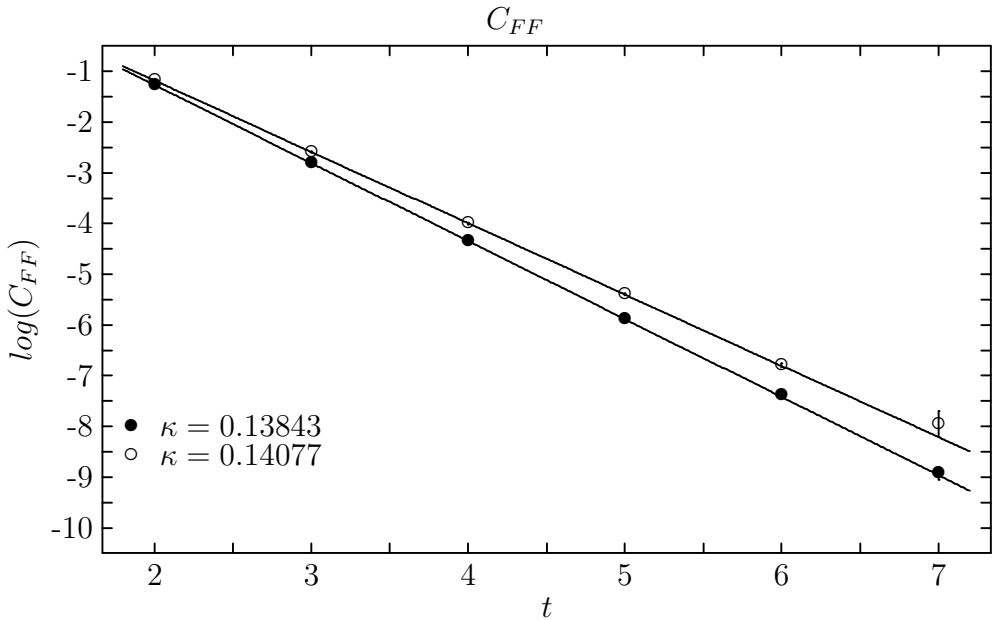}
\caption{Plot of $C_2^{FF}$ after the substraction for the contribution
of excited states. \label{figure1}}
\end{figure}

The non-perturbative strong interaction effects in spectator
contributions to inclusive decays are contained in the matrix elements
$M_i$ in eqs.~(\ref{eq:m1})--(\ref{eq:m4}). They have been evaluated
in a quenched simulation on a $12^3 \times 24$ lattice at $\beta=5.7$
using the SW improved action~\cite{sw},
\begin{equation}
S^{\mathrm{SW}}=S^{\mathrm{gauge}}+S^{\mathrm{Wilson}}-
\frac{i c_{\mathrm{SW}}}2 \sum_{x, \mu, \nu}
\overline{q} (x) F_{\mu \nu}(x)q(x),
\label{eq:action}\end{equation}
where $S^{\mathrm{gauge}}$ and $S^{\mathrm{Wilson}}$ are the Wilson
gauge and quark actions, respectively. We use 20 gauge-field
configurations and the light quark propagators are computed using a
stochastic inversion based on the exact relation
\begin{equation}
	(A^{-1})_{ij}=\frac1Z \int [d\phi] (A_{jk}\phi_k)^\ast
	\phi_i\ \exp \left( -\phi_l^\ast (A^\dagger A)_{lm} \phi_m \right)
\end{equation}
where the $\phi$ are auxiliary bosonic fields, introduced in order to
perform the inversion of the matrix $A$, which in our case is the
fermionic matrix. To reduce the statistical noise the technique of
maximal variance reduction has been used~\cite{cm1}.  The use of this
stochastic inversion technique makes it possible to compute a
light-quark propagator from each point in the half of the lattice with
$0<t\leq 12$ (we call this region box I) to each point in the other half
where $12<t\leq 24$ (box II). This increases considerably the effective
statistics in the computation of the matrix elements.

We have performed the calculations at $c_{\mathrm{SW}}=1.57$, which is
the numerical value of $1/u_0^3$, with $u_0$ being the the average
value of a link variable as defined from the trace of the
plaquette. The calculation is therefore ``tadpole-improved'' and hence
the perturbative coefficients in table~(\ref{table1}) are the same as
those which would be obtained with a tree-level improved action
($c_{\mathrm{SW}}=1$). We have evaluated the matrix elements with two
values of the light-quark mass, corresponding to $\kappa_1=0.13847$
(for which $m_\pi a\simeq 0.74(4)$) and $\kappa_2=0.14077$ (for which
$m_\pi a\simeq 0.52(3)$)~\cite{hps}. For this value of
$c_{\mathrm{SW}}$ $\kappa_{crit} \simeq 0.14351$ ~\cite{hps}.  The
same lattice has been used to compute, with satisfactory results, the
wave function of a $B$-meson and the effective coupling constant for
the decay $B^\ast \rightarrow B + \pi$ (in the Heavy Meson Chiral
Lagrangian). This computation is reported in ref.~\cite{me2} and the
values which we use for $f_B^2 m_B$ are extracted from this paper.

The evaluation of the matrix elements requires the computation of two-
and three-point correlation functions of the form,
\begin{equation}
C_2(t_x)=\sum_x \left< 0 \right| J(x) J^\dagger (0) \left| 0 \right>
\label{c2}\end{equation}
where we have assumed $t_x>0$, and
\begin{equation}
C_3({\cal O}, t_x,t_y)=\sum_{x,y} \left< 0 \right| J(y) {\cal O}(0)
J^\dagger (x) \left| 0 \right> 
\label{c3}
\end{equation}
where $t_y>0>t_x$.
In eqs. (\ref{c2}) and (\ref{c3}) $J$ and
$J^\dagger$ are interpolating operators which can destroy or create the
$\Lambda_b$ baryon, for which we take 
\begin{equation}
	J^\dagger_{\gamma} =
	\varepsilon _{abc} \left(
	\overline{u}_\alpha ^a\left( \gamma
	^5 C\right) _{\alpha \beta }\overline{d}_\beta^b\right) 
         \ \overline{b}_\gamma^c\ ,
\label{eq:jdef}\end{equation}
where $\overline{u}$, $\overline{d}$, $\overline{b}$ are the quark
fields.  In eq.(\ref{eq:jdef}) $a,\,b,\,c$ are colour labels,
$\alpha,\beta,\gamma$ are spinor labels and a sum over repeated
indices is implied.  We define $Z_\Lambda$ by
\begin{equation}
	Z_\Lambda u_\gamma^{(s)}({\bf 0}) =\frac{\left< \Lambda_b,s 
	\right| J^\dagger(0)_\gamma \left| 0
	\right>}{\sqrt{2 m_\Lambda}} 
\end{equation}
where $s$ represents the spin state (up or down) of the baryon. 

In order to enhance the contribution of the ground state to the
correlation functions, it is useful to ``smear'' the interpolating
operators $J$ and $J^\dagger$. In this paper we will follow
ref.~\cite{cm1} and adopt the type of smearing known as ``fuzzing''. 
This technique consists in replacing light quark field $q(x)$, by a
``fuzzed'' field
\begin{equation}
	q^F(x) = \sum_{i=1,2,3}
	U^F_i(x)\,q(x+\hat\i)+U^F_{-i}(x)\,q(x-\hat\i)\ , 
\end{equation}
where $U^F_{\pm i}(x)$ are defined by the recursive relations
\begin{align}
	U^F_i(x) ={\cal P}_{SU(2)} \Big[\zeta\, U^F_i(x) + \sum_{j\neq i}
	&U^F_j(x)U^F_i(x+\hat\j)U^F_{-j}(x+\hat\i+\hat\j)+\nonumber\\
	&U^F_{-j}(x)U^F_i(x-\hat\j)U^F_j(x+\hat\i-\hat\j) \Big]  \\
	U^F_{-i}(x) = {\cal P}_{SU(2)} \Big[\zeta \,
	U^F_{-i}(x) + \sum_{j\neq i}
	&U^F_j(x)U^F_{-i}(x+\hat\j)U^F_{-j}(x-\hat\i+\hat\j)+\nonumber\\
	&U^F_{-j}(x)U^F_{-i}(x-\hat\j)U^F_j(x-\hat\i-\hat\j) \Big]\ ,
\end{align}
starting with initial values $U^F_i(x)=U_i(x)$ and
$U^F_{-i}(x)=U^\dagger_i(x-\hat\i)$. ${\cal P}_{SU(2)}$ is a projector
on $SU(2)$, implemented as in the Cabibbo-Marinari cooling algorithm,
and $\zeta=2.5$ is a constant value. The recursive procedure for
$U^F_{\pm i}(x)$ has been applied twice.

We introduce two superscripts on each correlation function, each of
which can be either ``$F$'' or ``$L$'', which indicate whether the
interpolating operators $J$ and $J^\dagger$ are fuzzed or local.

The standard technique to extract hadronic matrix elements of the type
$\left< \Lambda_b\right| {\cal O} \left| \Lambda_b \right>$ is to look
for plateaus in the ratios
\begin{equation}
	R({\cal O}, t_1, t_2)=Z_\Lambda^2 \frac{C_3^{FF}({\cal O}, t_1,t_2)}
	{C_2^{LF}(t_1) C_2^{LF}(t_2)}\ .
\end{equation}
In our analysis, however, even with the use of fuzzed interpolating
operators $J$ and $J^\dagger$, we cannot eliminate the effects of
excited states from the three-point correlation functions
$C_3^{FF}({\cal O}, t_1,t_2)$ satisfactorily. On the other hand we do
find that the ground state dominates the two-point correlation
function $C_2^{FF}(t)$ for $t>3$, and the masses we obtain in this way
agree, within errors, with those found previously on the same
lattice~\cite{cm1} using a 3-mass correlated fit for a number of
smeared correlators. In order to obtain the matrix elements $M_i$ of
eqs.~(\ref{eq:m1})--(\ref{eq:m4}) we therefore need to subtract the
effects of the excited states.

\begin{figure}
	\input{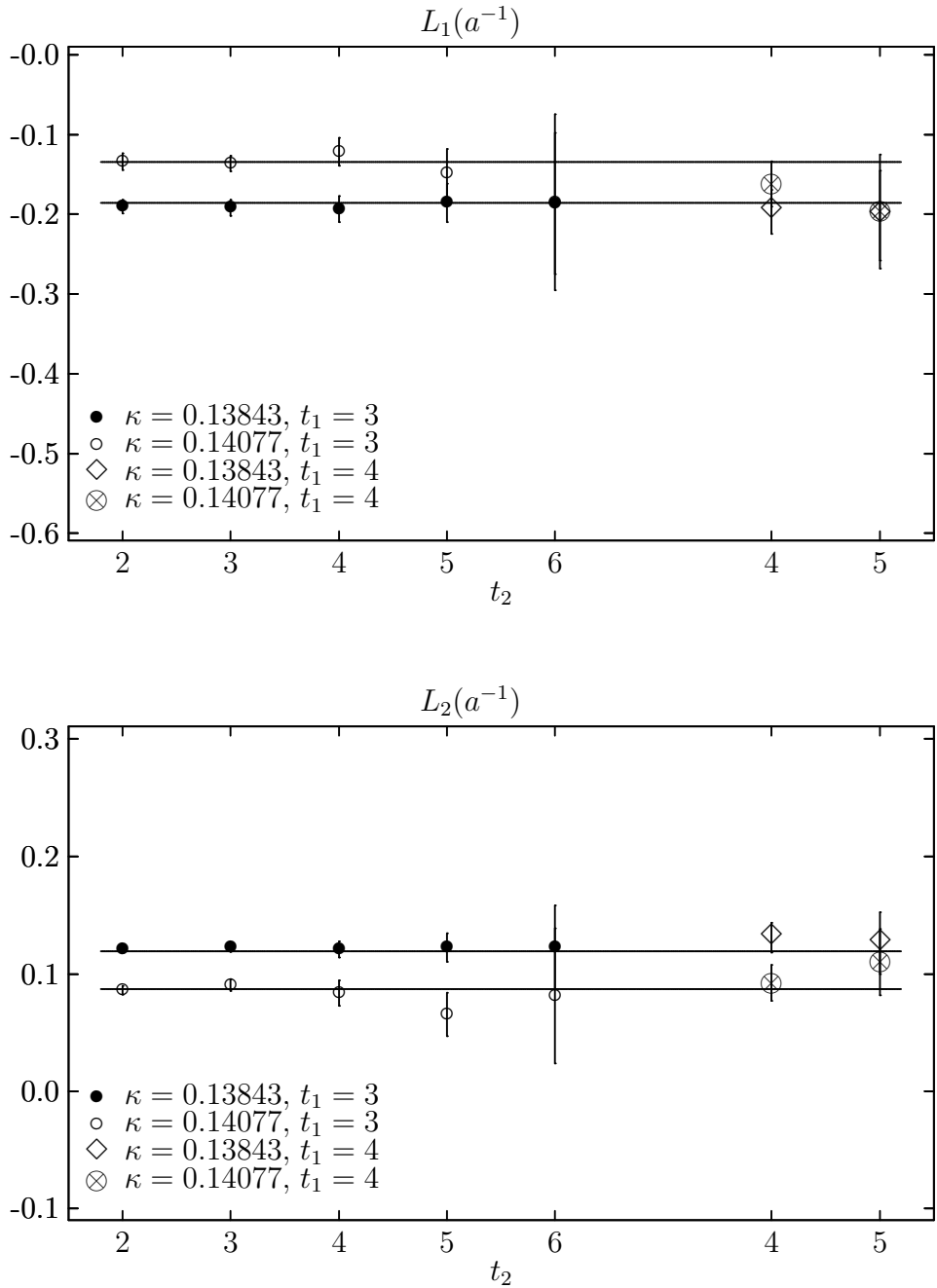}
\caption{Plot of the matrix elements $L_1(a^{-1})$ and $L_2(a^{-1})$ 
computed after the substraction of excited states. \label{figure2}}
\end{figure}

We have followed the following procedure to extract the matrix elements
$\left< \Lambda_b\right| {\cal O} \left| \Lambda_b \right>$:
\begin{itemize}
\item
For each value of the light-quark mass we start by fitting the two-point
correlation function for $t>3$ with a single exponential 
\begin{equation}
	C_2^{FF}(t)= (Z^F_{\Lambda_1})^2\exp(-m_\Lambda t)\ ,
\end{equation}
thus obtaining the mass of the ground state, $m_\Lambda$. Within
errors, the masses of the ground state which we obtain are in
agreement with those obtained from the same lattice using a more
sophisticated fitting procedure in ref.~\cite{cm1}.
\item 
We model the contribution of the excited states by a second exponential
and now fit $C_2^{FF}(t)$ for $t>1$ by 
\begin{equation}
	C_2^{FF}(t)=(Z^F_\Lambda)^2 e^{-m_\Lambda t}
	        +(Z^F_{\Lambda_1})^2 e^{-m_{\Lambda_1} t}\ ,
\end{equation}
keeping $m_\Lambda$ and $Z^F_\Lambda$ fixed at the values obtained from
the single exponential fit above. We find that $C_2^{FF}$ is well
represented by the two exponentials.

\item
For each operator ${\cal O}$ we then fit the three-point correlation
function\\ $C_3^{FF}({\cal O}, t_1, t_2)$ to
\begin{eqnarray}
	C^{FF}_3({\cal O},t_1,t_2)&=&\frac{\langle \Lambda_b | {\cal
			 O} | \Lambda_b 
			 \rangle}{2m_\Lambda}(Z^F_\Lambda)^2
			 e^{-m_\Lambda (|t_1|+t_2)} \\
			 &+& C \left[ 
			e^{-m_\Lambda |t_1| - m_{\Lambda_1} t_2}+ 
			e^{-m_\Lambda |t_1| - m_{\Lambda_1} t_2} \right]
\label{eq:cff3}\end{eqnarray}
obtaining values for the two unknown parameters $\left<
\Lambda_b\right| {\cal O} \left| \Lambda_b \right>$ and the constant
$C$, which encodes the contribution from excited states.
\end{itemize}
This procedure has been repeated for each of the 4 relevant operators,
and for the linear combinations corresponding to $L_1$ and $L_2$ on 40
jackknife samples to extract the statistical errors.

\begin{table}
\begin{center}
\setlength{\tabcolsep}{2pc}
\settowidth{\digitwidth}{\rm 0}
\catcode`?=\active \def?{\kern\digitwidth}
\begin{tabular*}{\textwidth}{@{}lrr} \hline
expression &$ \kappa _1$ & $\kappa _2$ \\ \hline 
$Z_A^2Z_L^2=f_B^2 m_B/2  $ & $  0.33(1)       $&$ 0.33(1)$ \\ 
$M_1 $ & $ -0.026(3) $&$ -0.019(3)$ \\
$M_2 $ & $ 0.045(4)  $&$ 0.039(5) $ \\
$M_3 $ & $ 0.018(2)  $&$ 0.013(2) $ \\
$M_4 $ & $ -0.040(4) $&$ -0.031(4)$ \\
$L_1(a^{-1}) $ & $ -0.18(2)  $&$ -0.13(3) $ \\ 
$L_2(a^{-1}) $ & $ 0.21(2)   $&$ 0.16(2)  $ \\ 
$L_1(m_b)    $ & $ -0.31(3)  $&$ -0.22(4) $ \\ 
$L_2(m_b)    $ & $ 0.23(2)   $&$ 0.17(2)  $ \\ \hline
\end{tabular*}
\end{center}
\caption{Lattice results for the matrix elements computed on lattice,
$M_i$, the combined matrix elements at two different scales, $L_i$,
and the physical ratio of lifetimes. \label{table2}}
\end{table}

In order to control the contributions from the excited states more
effectively it will be necessary to carry out a simulation with
considerably improved statistics. It is, however, possible to check
the consistency of our approach a posteriori. We subtract the
contributions from the excited states obtained above from the two- and
three-point correlation functions, and look for plateaus in the
ratios:
\begin{equation}
	R({\cal O}, t_1, t_2)=Z_\Lambda^2 
	\frac{\tilde C_3^{FF}({\cal O}, t_1,t_2)}
	{\tilde C_2^{LF}(t_1) \tilde C_2^{LF}(t_2)}
\label{eq:ratio}\end{equation}
where the tilde indicates that contribution from excited states has
been subtracted from the correlation function. The subtracted
two-point correlation function is reported in fig.~\ref{figure1}.
Fig.\ref{figure2} shows the plateaus for the ratios $R$ corresponding
to the operators in $L_1$ and $L_2$ (with the appropriate
normalization factor $\frac8{f_B^2 m_B}$). The plateaus in
fig.\ref{figure2} give us confidence in our treatment of the
subtraction of the contribution of the excited states. The results for
the matrix elements obtained from eqs.~(\ref{eq:cff3}) and
(\ref{eq:ratio}) agree to within 1\%.

Our results for the matrix elements at each of the two values of
$\kappa$ are reported in table~\ref{table2}.  Combining them with
eq.~(\ref{theo1}) we obtain
\begin{equation}
\frac{\tau (\Lambda _b)}{\tau (B^0)}=
	\left\{
	\begin{tabular}{ll}
	$0.91(1)$ & for $a m_\pi=0.74(4)$\\
	$0.93(1)$ & for $a m_\pi=0.52(3)$\ .
	\label{lambda_results}
	\end{tabular}
	\right. 
\end{equation}

>From eq.~(\ref{lambda_results}) we see that, although they are of
$O(1/m_b^3)$, spectator effects are indeed significant (compare
eqs.~(\ref{lambda_results}) and (\ref{theo1})). Estimates of the
parameter $r$ defined in eq.~(\ref{eq:rdef}), using the
non-relativistic quark model or the bag model~\cite{guberina,blok} or
QCD Sum Rules~\cite{qcdsr} are typically in the range 0.1--0.5. On the
other hand, Rosner has estimated $r$ from the spin splitting between
$\Sigma_Q$ and $\Sigma^*_Q$ baryons (Q=c,b) and finds $r\simeq 1$ (2)
from charmed (beauty) baryons~\cite{rosner}. A recent reanalysis of
this problem using QCD Sum Rules in which more condensates are
introduced, finds a range of possible values for the ratio of
lifetimes (including ones close to the experimental value in
eq.(~\ref{exp1})), depending on the (unknown) values of various
condensates~\cite{qcdsr2}.  At the two values of the light quark mass
at which we do our computations we find $r\simeq 1.2\pm 0.2$, which is
at the high end of expectations. The large values which we find for
$r$ and consequently the significant effect on the prediction for the
lifetime ratios, make it important to improve the precision of the
lattice simulations.

In quark models and related pictures, the parameter $\tilde B=1$. In our
simulations we find larger values, $\tilde B=1.9\pm 0.2$ ($1.3\pm 0.2$)
at $\kappa_1$ ($\kappa_2$).

\section{Conclusions}

In this paper we have evaluated the matrix elements which contain the
non-pertubative QCD effects in the spectator contribution to inclusive
decays of the $\Lambda_b$ baryon. Our principal results (in the
$\overline{\text{MS}}$ scheme) are presented in table~\ref{table2}. The
results indicate that spectator effects are important, accounting for a
significant fraction of the discrepency between the theoretical
prediction for $\tau(\Lambda_b)/\tau(B_d)$ in eq.~(\ref{theo1}) and the
experimental result in eq.~(\ref{exp1}). It also appears that not all of
the discrepency can be accounted for  by spectator effects.

The calculation described in this paper is the first evaluation, using
lattice simulations, of the matrix elements in eq.~(\ref{elle1}) and
eq.~(\ref{elle2}) between $\left| \Lambda_b \right>$ states. Having
established that spectator effects are significant, it is now
necessary to improve the precision, both statistical and
systematic. This requires a high-statistics simulation at a smaller
value of the lattice spacing (to decrease the errors due to
discretisation) and with more values of the light quark mass (to
enable a reliable extrapolation to the chiral limit). 

In this study we have used static $b$-quarks. It would also be valuable,
as a control of the systematic errors, to repeat the calculation with
propagating $b$-quarks, for which these uncertainties are different.

\section*{Acknowledgements}

We wish to acknowledge Giulia De Divitiis, Luigi Del Debbio, Jonathan
Flynn, Hartmut Wittig and other colleagues from the UKQCD
collaboration for helpful discussions. Our simulations have been
performed on a Cray J90 at the Rutherford Appleton Laboratory
(Oxford).

This work was supported by PPARC grants GR/L29927 and GR/L56329, and
EPSRC grant GR/K41663.

\end{document}